# Fat-free noncontrast whole-heart CMR with fast and power-optimized off-resonant water excitation pulses

Adèle L.C. Mackowiak[1,2,3], Davide Piccini[3,4], Jessica A.M. Bastiaansen[1,2]

[1] Department of Diagnostic, Interventional and Pediatric Radiology (DIPR), Inselspital, Bern University Hospital, University of Bern, Switzerland

[2] Translation Imaging Center (TIC), Swiss Institute for Translational and Entrepreneurial Medicine, Bern, Switzerland

[3] Department of Radiology, Lausanne University Hospital (CHUV) and University of Lausanne (UNIL), Lausanne, Switzerland

[4] Advanced Clinical Imaging Technology (ACIT), Siemens Healthcare AG, Lausanne, Switzerland

**Corresponding author**: Jessica A.M. Bastiaansen, PhD, Assist. Prof.

E-mail: jbastiaansen.mri@gmail.com

Department of Diagnostic, Interventional and Pediatric Radiology

Inselspital, Bern University Hospital, University of Bern, Switzerland

Freiburgstrasse 3, 3010 Bern, Switzerland

# ABSTRACT

Background

Cardiovascular MRI (CMR) faces challenges due to the interference of bright fat signals in visualizing structures like coronary arteries. Effective fat suppression is crucial, especially when using whole-heart CMR techniques. Conventional methods often fall short due to rapid fat signal recovery, resulting in images with residual fat content. Water-selective off-resonant RF pulses have been proposed but come with tradeoffs between RF pulse duration, which increases scan time, and increased RF energy deposit, which limits their applicability due to SAR constraints. The study introduces a lipid-insensitive binomial off-resonant (LIBOR) RF pulse, which addresses concerns about SAR and scan time, and aims to provide a comprehensive quantitative comparison with published off-resonant RF pulses for CMR at 3T.

Methods

A short (1ms) LIBOR pulse, with reduced RF power requirements, was developed and implemented in a free-breathing respiratory self-navigated 3D radial whole-heart CMR sequence at 3T. A BORR pulse with matched duration, as well as previously published LIBRE pulses (1ms and 2.2ms), were implemented and optimized for fat suppression in numerical simulations and validated in healthy subjects (n=3). Whole-heart CMR was performed in healthy subjects (n=5) with all four pulses in randomized order. The signal-to-noise ratio (SNR) of ventricular blood, skeletal muscle, myocardium, and sub-cutaneous fat, and the coronary vessel sharpness and length were compared.

Results

Experimental results validated numerical findings and near homogeneous fat suppression was achieved with each of the four pulses. Comparing the short RF pulses (1ms), LIBOR reduced the RF power nearly two-fold compared with LIBRE, and three-fold compared with BORR, and LIBOR significantly decreased overall fat SNR. The reduction in RF pulse duration (from 2.2ms to 1ms) shortened the whole-heart acquisition from 8.5min to 7min. No significant differences in coronary arteries detection and sharpness were found when comparing all four pulses.

Conclusion

The use of LIBOR pulses enabled whole-heart CMR under 7 minutes at 3T, with large volume fat signal suppression, while reducing RF power compared with LIBRE and BORR pulses. LIBOR is an excellent candidate to address SAR problems encountered in CMR sequences where fat suppression remains challenging and short RF pulses are required.

# INTRODUCTION

In cardiac MRI (CMR), bright fat signals hinder the visualization of small anatomical structures such as coronary arteries, which are embedded in adipose tissue[1]. When fat signal is not sufficiently suppressed, it can generate India ink artifacts contouring arteries[2] and can degrade overall image quality by the introduction of streaking artifacts[3], when sub-cutaneous chest fat is not sufficiently suppressed. Fat signal suppression is thus essential in CMR, but especially challenging when using advanced whole-heart CMR techniques that employ non-Cartesian readout strategies[4,5].

The rapid T1 recovery of the fat signal, following the application of conventional fat saturation, contributes to significant fat signal weighting in the k-space center of 3D radial acquisitions, which results in cardiac images where fat remains largely present[3], albeit with lower signal intensity. Therefore, acquisition strategies are preferred that uniquely measure the water signal, either by using dedicated radiofrequency (RF) pulses for water selective excitation[6], or by fast interruptions of the bSSFP steady-state[7–9].

Commonly used RF pulses for water excitation follow a binomial pattern, in which the spacing between the two subpulses is used to create a different excitation profile for both water and fat, which makes them time-inefficient, with typical pulse durations on the order of 2 to 3ms at 3T. By using a phase modulation on the second RF subpulse[10–13], as for example done in a 1-90°-1 water excitation pulse, the duration can be reduced by half compared with a conventional 1-180°-1 water excitation pulse, at the expense of fat suppression bandwidth.

Excellent fat suppression capabilities have been reported for MR angiography when using a pair of binomial off-resonant rectangular (BORR) pulses with opposing phase. The BORR pulse[14,15], developed by Ye et al., uses a 180° phase modulation on the second subpulse combined with a change of the RF excitation frequency. Although theoretically promising because of the wide fat suppression band it offers, a reduction of the BORR pulse duration below 2.6ms has not been investigated.

The development of a lipid insensitive binomial off-resonant RF excitation (LIBRE) pulse offered reduced pulse durations while outperforming the fat suppressing capabilities of regular fat suppression methods and water excitation approaches[16]. The duration of the LIBRE pulses can be as short as 1ms at 3T[17], and they have been used for 3D radial whole-heart CMR at both 1.5T[18] and 3T[3], including free-running 5D acquisitions[18,19]. LIBRE pulses were successfully employed in coronary MR angiography in the clinical setting[19–21], enabling the detection of coronary aneurysms in children with Kawasaki disease at 3T[21]. Besides CMR, the pulses added value for musculoskeletal[17,22], brain[23] and moving eye[24] imaging applications.

To achieve shorter scanning times, off-resonant pulse durations can be decreased by increasing their RF excitation frequency, which makes the RF pulse less efficient in exciting on-resonant water. To mitigate this efficiency loss, the RF power needs to be increased to achieve the desired water signal. Therefore, a decrease in pulse duration comes at the expense of increased RF power and RF energy deposits (SAR). This poses practical limits on the use and flexibility of off-resonant excitation pulses where short repetition times (TR) are required, especially when combined with already SAR intensive acquisitions like bSSFP[18]. No attempts have been made to optimize off-resonant water excitation pulses in terms of RF power deposition, nor has a comprehensive comparison been conducted between the various off-resonant water excitation pulses published to date.

The aim of the study was to implement a novel lipid-insensitive binomial off-resonant (LIBOR) RF pulse for water excitation. The LIBOR pulse was designed to address concerns on RF energy deposit and total pulse duration, and was implemented for noncontrast free-breathing respiratory self-navigated whole heart CMR at 3T. With the additional implementation of LIBRE and BORR pulses of equivalent pulse duration (1ms), a comprehensive quantitative comparison was provided between the different types of binomial off-resonant pulses for water excitation. Their fat suppression capabilities were quantified using numerical simulations and in healthy subjects.

## METHODS

### RF pulse design of LIBOR

LIBOR was designed as a phase-modulated off-resonant water-excitation pulse with a total RF duration of 1ms. To increase water excitation efficiency of the pulse, the LIBOR excitation frequency was halved to 780Hz compared with a LIBRE pulse of 1ms duration[17]. The lowering of the excitation frequency makes the LIBOR pulse by design more efficient in exciting water and thus requires a lower RF power. Then, similarly to phase modulation of a 1-90°-1 water excitation pulse[12], the second LIBOR sub-pulse requires a phase offset to achieve fat signal suppression, which was determined using numerical simulations of the Bloch equations, which were additionally verified in MRI experiments. Simulations were performed to compute the transverse magnetization $M_{xy}$ as function of phase-modulation and off-resonance after a single LIBOR RF excitation with an RF excitation angle of 10°. Off-resonance was varied between - 800Hz and 800Hz in steps of 5Hz and the phase-modulation between 0° and 360° in steps of 5°. The phase modulation corresponding to the widest signal suppression band around fat, defined as 10% of the maximum observed transverse magnetization ($M_{xy}$), was selected for subsequent measurements. All simulations were performed in MATLAB 2021 (The MathWorks, Natick, MA, USA). The code is shared on our github page.

## Numerical simulations of LIBOR, LIBRE and BORR pulses

Additional simulations were performed as described earlier[3,16] to compute the transverse magnetization ($M_{xy}$) of GRE sequences using a LIBOR, BORR, and LIBRE pulse of 1ms duration. An additional LIBRE pulse of 2.2ms duration was added for the comparison, because this pulse has been investigated before for whole-heart radial CMR at 3T[3], in which it was evaluated against conventional fat suppression and water excitation. The $M_{xy}$ was determined as function of RF excitation angle and off-resonance, providing a quantitative comparison of water excitation efficiency, i.e., the required increase in power deposit relative to on-resonant pulses, fat signal suppression, and corresponding suppression bandwidths. RF excitation angles were varied between 2° and 30° in steps of 2°, the off-resonance was varied between -800Hz and 800Hz in steps of 5Hz, and a repetition time (TR) of 5ms, a $T_1$ of 2000ms, and a $T_2$ of 50ms were used. To ensure steady state, 500 excitations were simulated with perfect RF and gradient spoiling by nulling $M_{xy}$ after each RF excitation. The results of these simulations informed on RF power increases for subsequent MRI experiments. The RF waveforms corresponding to each off-resonant pulse were plotted for comparison.

## MRI experiments

Volunteer experiments were performed on a 3T clinical MRI scanner (MAGNETOM Prisma[fit], Siemens Healthcare, Erlangen, Germany) after obtaining written and informed consent from all participants. The study was approved by a local Ethics Committee (authorization CER-VD 2021-00708, Lausanne, Switzerland).

A LIBOR, BORR, and LIBRE RF pulse were implemented in a 3D radial spoiled GRE sequence following a spiral phyllotaxis trajectory[4], with the option to perform ECG-triggered and respiratory-self-navigated whole-heart acquisitions, as described in[3,5]. The user interface of the scanner console was modified to select between the LIBOR, BORR and LIBRE RF pulses by varying the duration (τ) of the sub-pulses, their RF excitation frequency ($f_{RF}$), and the phase modulation of the second sub-pulse (ΔΦ) (**Table 1**). The research sequence can be requested on the exchange platform (TeamPlay) of the vendor.

Sequence parameters included: an isotropic field-of-view (FOV) of (200mm)$^3$, a spatial resolution of (1.14mm)$^3$, and a pixel bandwidth of 888Hz/px. For the scans using LIBOR, BORR, and LIBRE RF pulses with a total duration of 1ms, the echo time and repetition time were TE/TR=1.99ms/4.30ms. For LIBRE scans with a total pulse duration of 2.2ms, the TE/TR was 2.59ms/5.53ms. The RF excitation angles varied for each pulse because of their different RF power requirements (**Table 1**). Image reconstruction for all experiments was performed at the scanner, and the SAR was recorded.

**RF pulse optimization in vivo**

Knee experiments were performed to test the effect of varying the RF excitation frequency (LIBRE, BORR) or RF phase modulation (LIBOR) on the measured water and fat signals. The knees of three subjects were scanned using a 35-channel knee coil array. The 3D radial trajectory consisted of 513 spiral segments, which are successively rotated by the golden angle about the longitudinal axis. Each spiral segment was made of 24 readout lines that go from one periphery of the k-space to the opposite, while passing through the k-space center. This trajectory has been previously described mathematically and has been shown to provide a general reduction of eddy current artifacts and improvement in image quality[4].

The phase offset of LIBOR was varied from 270° to 340° in steps of 10°, the frequency $f_{RF}$ of BORR and LIBRE (1ms) was varied from 1500Hz to 1700Hz in steps of 20Hz, and the frequency $f_{RF}$ of LIBRE (2.2ms) was varied from 400Hz to 560Hz in steps of 20Hz.

The acquisition time was TA=0:53min per scan using LIBOR, BORR and LIBRE (1ms) pulses, and TA=1:08min per scan with LIBRE (2.2ms) and was fixed across subjects.

**Noncontrast free-breathing respiratory self-navigated whole-heart MRI at 3T**

Free-breathing ECG-triggered respiratory self-navigated whole-heart MRI acquisitions were performed with 100% scan efficiency in five subjects using an 18-channel chest coil array. A 3D radial spiral phyllotaxis pattern described in the previous section was used. A diastolic window for data collection was maintained at 100ms per heartbeat. For the LIBOR, BORR and LIBRE scans with a 1ms total pulse duration, the acquisition was segmented into 424 spirals, each comprising of 23 lines. For the LIBRE scans with the 2.2ms total RF pulse duration, the trajectory was 18 lines and 547 spirals. This provided a matching 3D spiral pattern while keeping the same acquisition window for all scans, including those using a longer RF pulse. A vendor-provided T2-preparation module[25] of 40ms was used to improve blood-myocardium contrast, which was played out with each heartbeat. The total amount of acquired k-space lines was the same for all scans (~10k). Respiratory-motion-corrected cardiac images were reconstructed at the scanner, using a method for respiratory self-navigation[26] based on blood pool identification and signal intensity variations in one-dimensional superior-inferior projections[5].

## Image analysis

The DICOM images reconstructed at the scanner were analyzed by computing the signal-to-noise ratio (SNR) and contrast-to-noise ratio (CNR) in manually drawn regions of interest (ROIs) using the ImageJ software[27]. The brightness and contrast of all images were automatically matched in the software. ImageJ provided the area, mean, and standard deviation of the signal intensity measured in the ROIs.

The following definitions of SNR and CNR were used for all analyses presented in this study.

$$\text{SNR}_A = \frac{mean(S_A)}{SD(S_{bkg})} \quad\quad\quad [\text{Eq.1}]$$

$$\text{CNR}_{A-B} = \text{SNR}_A - \text{SNR}_B \quad\quad\quad [\text{Eq.2}]$$

where $S_A$ is the signal intensity of tissue A, and $S_{bkg}$ is the signal intensity in the background of the image.

**Knee MRI**

ROIs corresponding to three tissue types (bone marrow, sub-cutaneous fat and *vastus medialis* muscle) were drawn in 5 slices of one of the four 3D datasets, and re-used to analyze data from the other three water excitation pulses. The shape of the ROIs was only adjusted if visible motion between the separate acquisitions was detected, and the area was kept as similar as possible. An additional ROI was drawn in the background of each image. The order of data analysis was randomized. SNR were reported as a mean and standard deviation across all subjects. For each water excitation pulse, the tuning parameter (RF excitation frequency or phase) that maximizes muscle-fat CNR (defined according to [Eq.2]) was chosen as the optimal RF parameter for that pulse and used in subsequent cardiac experiments.

**Whole-heart MRI**

ROIs placed in the left-ventricular blood pool, in the myocardium muscle, and in the chest sub-cutaneous fat, were drawn in each volunteer. The ROIs were used to analyze the datasets obtained with the four water excitation pulses, with adjustments in shape where necessary, and a randomized order of. The SNR of the three tissue types, as well as the blood-fat CNR, were reported as mean and standard deviation across all subjects.

All 3D whole-heart imaging volumes were analyzed for the detection and visualization of the right coronary artery (RCA), the left main (LM) and the left anterior descending (LAD) coronary arteries using the MR angiography analysis software SoapBubble[28]. Coronary reformats were obtained by tracking the vessels of interest. Vessel detection rates, as well as vessel sharpness along the first 4 cm when applicable, were reported.

**Statistical methods**

All results are presented as mean ± standard deviation of the mean (SD). For SNR and CNR analyses, as well as for coronary sharpness analysis, statistically significant differences between the proposed LIBOR pulse and the three other water excitation pulses were assessed using paired two-tailed Student's t-tests, corrected for multiple comparisons (GraphPad Prism, San Diego, CA, USA). In all analyses, p<.05 was considered statistically significant.

# RESULTS

## Numerical simulation experiments

Simulations experiments show excellent fat suppression with a bandwidth of 300Hz of the novel LIBOR pulse using a -45° degrees phase modulation (**Fig.1**), in comparison to a bandwidth of 130Hz of a conventional 1-90°-1 pulse (data not shown). The simulated GRE signal attests of the different RF power requirements in terms of RF excitation angle increase (**Fig.2**) and demonstrates similar B1 behavior of the different pulses. The LIBOR water excitation efficiency is higher compared with the LIBRE and BORR pulses of equivalent duration, and required the lowest RF power increase compared with a LIBRE pulse of 2.2ms. Although the BORR pulse achieved the widest suppression band (**Fig.2**), it required the highest RF power increase.

## 3D radial MRI in the knee for RF pulse calibration and comparisons at 3T

A phase modulation of ΔΦ=320° (or -40°) for the LIBOR pulse achieved lowest fat SNR and highest muscle SNR (**Fig.3**), matching numerical findings. Shortening the BORR pulse to 1ms was possible and the lowest fat SNR was obtained using an RF excitation frequency of ~1540Hz (**Table 1**). Using optimized RF parameters for each pulse, the average SAR was significantly reduced when using the LIBOR RF pulse (**Table 1**).

Qualitatively, homogeneous fat signal suppression was achieved in subcutaneous fat and bone marrow with all RF pulses, resulting in good visualization of the various muscles and cartilages (**Fig.4**). Images obtained with BORR displayed signal bleed at some of the air-tissue interfaces (**Fig.4**, green arrows). Minor amounts of signal bleed were also observed in the LIBOR (1ms) case, and may be attributed to a combination of off-resonance effects and power increase.

Quantitatively, in bone marrow, a significant signal decrease was detected between LIBOR (7.2±0.4) and the 2.2ms-long LIBRE pulse (8.8±0.9, p<.0001). In sub-cutaneous fat, significant SNR decreases were detected between LIBOR (16.5±1.1) and LIBRE (1ms) (19.2±2.0, p<.0001) and between LIBOR and LIBRE (2.2ms) (24.3±1.5, p<.0001). The only significant difference in SNR between LIBOR and BORR was detected in the muscle (82.8±14.1 vs 52.4±12.2, p<.00001). The LIBOR pulses also yielded a different muscle SNR than LIBRE (1ms) and LIBRE (2.2ms) (**Fig.5**). The highest average muscle-fat CNR across subjects was (66.3±13.8) for LIBOR, (54.0±14.0) for LIBRE (1ms), (34.4±12.4) for BORR, and (61.1±16.6) for LIBRE (2.2ms), obtained with the respective optimal RF tuning parameters as follows: ΔΦ=320° for LIBOR, $f_{RF}$=1620Hz for LIBRE (1ms), $f_{RF}$=1540Hz for BORR and $f_{RF}$=520Hz for LIBRE (2.2ms) (**Table 1**).

## Noncontrast respiratory self-navigated whole-heart MRI at 3T

Free-breathing whole-heart MRI was performed with 100% scan efficiency in all volunteers in under 7 minutes. Acquisition times varied based on the subject's heart rate (**Table 2**) but were on average reduced by 20% using the 1ms RF pulses. Recorded SAR values were lowest using the LIBOR pulse, when comparing the other RF pulses of short duration (**Table 2**).

Visually, all pulses performed well for fat suppression (**Fig.6A**), with LIBOR demonstrating more homogenous fat suppression of subcutaneous chest fat (**Fig.6A**, blue arrows). Overall, the SNR measurements on LIBOR whole-heart images were comparable to that of the other 1ms RF pulses, but consistent significant differences were found when comparing it to the 2.2ms LIBRE pulse (**Fig.6B**). LIBOR provided the lowest sub-cutaneous fat SNR (**Fig.6B**). LIBOR provided the highest blood-fat contrast measurement, with a value comparable with that of the short 1ms LIBRE pulse (Fig.6C), but significantly higher than both BORR (p<.0001) and 2.2ms-LIBRE pulses (p<.001).

The coronary arteries could be visualized using all four water-excitation pulses (**Fig.6**), with matching detection rates in all cases, with the RCA being detected in all subjects, the LM and LAD being detected in the same 4/5 subjects and the LCX being detected in the same 3/5 subjects (**Table 2**). In the subject where only the RCA was detected, unsuccessful respiratory-self-navigation caused residual respiratory motion artifacts, leading to a decreased image quality and difficulties to produce coronary reformats. This, however, affected all the scans of this subject equally and therefore did not affect the comparison between the four water-excitation pulses.

Only minor differences could be observed in the coronary reformats, and the overall visualization of the coronaries was similar with all tested pulses, leading to the same expected edge and location of the vessel (**Fig.6**). While LIBOR provided higher vessel sharpness on the first 4cm of the RCA than BORR, LIBRE (1ms) and LIBRE (2.2ms), statistical comparisons between LIBOR and the other three water-excitation pulses were all found to be non-significant (**Tab.2**).

## DISCUSSION

In this study, a novel off-resonant RF pulse for water excitation, LIBOR, was designed and optimized for noncontrast whole-heart CMR at 3T. Effective and fast lipid signal suppression over large volumes was obtained with reduced RF power and thus reduced SAR deposits.

In comparison with published off-resonant RF pulse designs such as BORR[14] and LIBRE[16], the proposed LIBOR technique utilizes a phase modulation on the second subpulse to reduce signal coming from fatty tissues. By design, the different RF pulses appear similar, but the LIBOR RF excitation frequency is halved compared with LIBRE and BORR, which reduces the required RF power to excite water to the same extent as the other RF pulses. The optimal LIBOR phase modulation was found through numerical simulations and was experimentally verified.

A characteristic feature of off-resonance water excitation pulses is their flexible pulse duration[29], which can be reduced at the expense of RF power. In the present study, it was observed that when the durations of the LIBRE, BORR, and LIBOR pulses were reduced to 1ms, the RF power was highest for the BORR pulse and lowest for the LIBOR pulse. Interestingly, the BORR pulse was not the most off-resonant in terms of excitation frequency. This highlights that the behavior of RF pulses is not always straightforward, and even if RF pulse designs appear similar, their performances can differ. The anticipated decrease in RF power and SAR deposits with the LIBOR pulse was validated by in vivo experiments. It is important to note that the SAR values recorded during cardiac MRI experiments were predominantly influenced by the power-demanding T2 preparation module used to generate blood-to-myocardium contrast in the T1-weighted GRE sequence[25].

Although the BORR pulse had the highest RF power requirement, it offered the widest fat suppression band. So far, it was not investigated whether a shortening of the BORR pulse duration was possible. Prior work utilized BORR pulses with durations of 2.6ms and a RF excitation frequency of 300Hz[14,15]. With these published settings, the RF power increases are minimal. This observation suggests that the use of BORR pulses could benefit acquisitions where scan time is less important and where TR can thus be increased. In this case the BORR pulse may offer great performance in terms of fat signal suppression and water excitation. For sequences requiring a short TR, BORR would be a less ideal candidate. This is strengthened by observations of signal smearing in vivo when a 1ms BORR pulse was used, likely caused by the high RF power requirement in combination with off-resonance effects. Similar signal bleeds, albeit to a lesser extent, were observed for the 1ms LIBRE pulse.

Experiments on the human knee corroborated findings from numerical simulations and were utilized to select the optimal parameters for each type of RF pulse, ensuring a fair comparison among them. However, small deviations in RF pulse parameters have minimal impact on the final results. In all in vivo experiments, the parameters for the LIBRE pulses were consistent with those identified in previous studies conducted years ago on the same MRI scanner (LIBRE 1ms[17], LIBRE 2.2ms[3,16]). This underscores that RF pulses do not necessitate dedicated fine-tuning when implemented and operated on different MRI scanners. Such fine-tuning was not necessary when a version of the whole-heart CMR sequence that uses LIBRE pulses[3] was installed and utilized at different clinical sites[19–21].

For whole-heart CMR, the use of short water excitation pulses of 1ms reduced the average scan time significantly, from 8:29 minutes to 6:50 minutes, while neither image quality nor coronary artery visualization were significantly affected. Detection rates and observed vessel sharpness were similar for all tested RF excitation pulses, as expected. Although the theoretical fat suppression bands varied for different RF pulses tested in the current study, fat signal suppression was visually similar in all images. The study did not utilize advanced motion compensation or image reconstruction techniques such as compressed sensing, and images were directly reconstructed at the scanner with an inline respiratory motion correction, provided by the vendor[5]. Therefore, the cardiac and coronary image quality could potentially be improved when using advanced image processing methods. To facilitate research collaborations in this direction, the anonymized raw data, as well as code to read it, is made available to the research community in a public repository.

The benefits of using off-resonant water excitation pulses in 3D radial free-breathing respiratory self-navigated whole-heart MRI have been reported in our prior work[3], notably within the frame of a comparison between 2.2ms LIBRE pulses and conventional fat suppression methods. In this prior work, LIBRE outperformed conventional methods and significantly improved visualization of coronary arteries, thanks to improved fat signal suppression, which in turn had a beneficial effect on respiratory self-navigation and helped reduce streaking artifacts. Therefore, the current study focused on a comprehensive quantitative comparison between different off-resonant RF excitation pulses.

The LIBOR pulse significantly reduces RF power deposits and offers a short TR, making it potentially ideal for acquisitions where a short TR is essential and where the acquisition is SAR intensive, such as with balanced steady-state free-precession (bSSFP) sequences. Offering a short TR becomes even more important when lowering the magnetic field strengths, for example to 1.5T and below. At lower magnetic field strengths, water and fat have smaller frequency differences compared with 3T, which typically translates into an increased duration of water excitation pulses. For example, to suppress fat signals with acceptable TR during whole-heart bSSFP acquisitions at 1.5T using LIBRE, a nominal RF excitation angle of approximately 120 degrees was chosen[18]. However, this choice was influenced by the imposed SAR limit, which limited our ability to achieve the optimal water excitation and, consequently, the water signal-to-noise ratio (SNR). Utilizing LIBOR RF pulses for whole-heart CMR at 1.5T, or at even lower magnetic field strengths such as 0.55T, could be a promising approach to address SAR issues and challenges in fat signal suppression. A comparative study between off-resonant RF pulses and various methods for fat signal suppression in whole-heart acquisitions, such as fast-interrupted steady-state sequences[7–9], the application of fat-suppressing T2 preparation modules[30,31], or whole-heart water fat separation[32] techniques is a topic of ongoing research[33].

Homogenous fat signal suppression is challenging in large volumes, and even more so for non-Cartesian whole-heart MRI acquisitions. A comprehensive comparison was made between different off-resonant RF water excitation pulses for their use in CMR at 3T. This included the development of a novel LIBOR pulse and the shortening of the BORR pulse. The successful implementation and comparison of LIBOR, BORR and LIBRE pulses in a 3D radial sequence for noncontrast respiratory self-navigated whole-heart MRI reveals that LIBOR is the most promising off-resonant water excitation pulse for fat signal suppression in large volumes with reduced SAR.

## CONCLUSION

A novel LIBOR RF pulse was developed for fast water excitation, and was implemented in a 3D radial phyllotaxis GRE sequence enabling a noncontrast free-breathing respiratory self-navigated whole-heart MRI acquisition in under 7 minutes. LIBOR demonstrated homogeneous fat suppression while reducing RF power and SAR compared with off-resonant pulses such as LIBRE and BORR. These findings are especially interesting to address SAR problems encountered in MRI sequences where fat suppression remains challenging.

## ACKNOWLEDGMENTS

This study was supported by funding received (by J.A.M.B.) from the Swiss National Science Foundation (grants #PCEFP2_194296, #PZ00P3_67871), the University of Lausanne Bourse Pro-Femmes, the Emma Muschamp Foundation and the Swiss Heart Foundation (grant #FF18054).

## AVAILABILITY OF DATA AND MATERIALS

An online repository containing the anonymized human MRI raw data, as well as RF pulse shapes used in this study is publicly available at: https://zenodo.org/records/8338079

Matlab code to 1) simulate the different RF pulses within a GRE sequence and 2) to read and display the anonymized raw data is available from: https://github.com/QIS-MRI/LIBOR_LIBRE_BORR_SimulationCode

The compiled research sequence can be requested through the Teamplay platform of Siemens Healthineers.

## CONTRIBUTIONS

AM and JB designed the study, performed the experiments, analyzed the data and wrote the main draft of the manuscript. JB developed and implemented the LIBRE RF pulses, DP contributed the source code for the respiratory-self-navigated CMR sequence. All authors read, revised, and approved the final manuscript.

## DISCLOSURE

Davide Piccini is an employee of Siemens Healthcare AG.

# TABLES & FIGURES

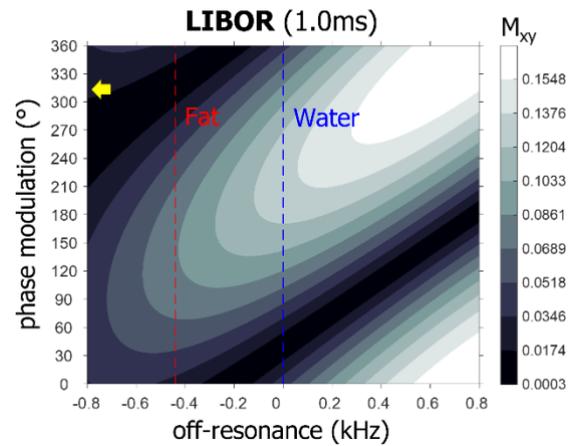

## Figure 1 – Frequency response profile of LIBOR

Frequency response profile of the LIBOR pulse as function of phase modulation applied to the second RF subpulse. Results were obtained following a numerical simulation of the transverse magnetization $M_{xy}$ after a single LIBOR RF excitation of $\alpha=10°$. These results indicate that a phase modulation of about 315°(yellow arrow), equivalent to -45°, results in a fat suppression band around the expected fat resonance at -440Hz.

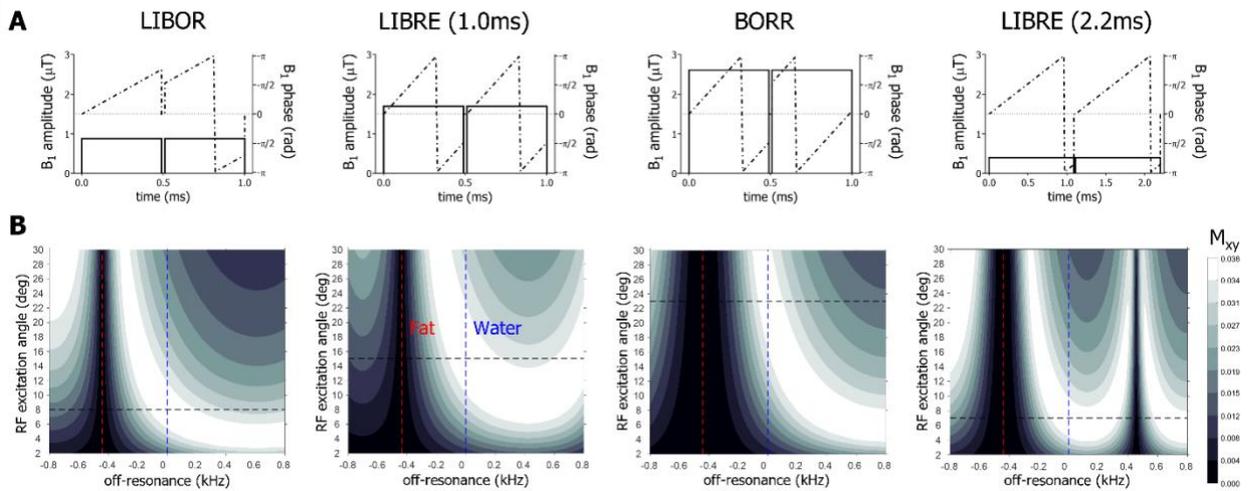

Figure 2 – RF pulse shape and magnetization response

(A) RF pulse waveforms illustrating the $B_1$ amplitude (solid line) and $B_1$ phase (dashed line) for the LIBOR, LIBRE, BORR, and LIBRE (2.2ms) pulse for maximized water excitation, which corresponds to the maximum observed $M_{xy}$ for water (0Hz) indicated in panel (B).

(B) Bloch simulations of a GRE sequence with the four different off-resonant pulses illustrate the effect on the transverse magnetization $M_{xy}$ as function of RF excitation angle and off-resonance (500 excitations, TR=5ms, $T_1$=2000ms, $T_2$=50ms). These simulations depict the expected water (0Hz, blue dashed line) and fat (-440Hz, red dashed line) signal behavior, including the optimal RF excitation angle for water excitation (intersection of the black and blue dashed lines), which provides a relative measure of the required RF power in comparison with the other RF pulses. The intersection of the dashed red and black lines provides an indication of the fat suppression band. The optimal water $M_{xy}$ for each pulse corresponds with the $B_1$ amplitude in panel (A). The BORR pulse shows the widest fat suppression band but requires the highest B1 amplitude.

| MR parameters | LIBOR (1.0ms) | LIBRE (1.0ms) | BORR (1.0ms) | LIBRE (2.2ms) |
|---|---|---|---|---|
| Excitation frequency $f_{RF}$ (Hz) | 780 | $f_{RF} = 1620$ | $f_{RF} = 1540$ | $f_{RF} = 520$ |
| Subpulse duration τ (μs) | 500 | 500 | 500 | 1100 |
| Phases $\Phi_1$, $\Phi_2$ | $\Phi_2 = \Phi_1 + \Delta\Phi$ $\Delta\Phi = 320°$ | $\Phi_2 = \Phi_1 + 2\pi\tau f_{RF}$ | $\Phi_2 = \Phi_1 + \pi$ | $\Phi_2 = \Phi_1 + 2\pi\tau f_{RF}$ |
| **Knee experiments** | | | | |
| RF excitation angle (deg) | 12 | 23 | 35 | 10 |
| SAR (mW.kg$^{-1}$) | 2.07 | 9.18 | 21.3 | 0.62 |
| **Cardiac experiments** | | | | |
| RF excitation angle (deg) | 19 | 38 | 56 | 16 |
| SAR (W.kg$^{-1}$) | 0.59 | 0.65 | 0.74 | 0.58 |

## Table 1 – RF pulses properties with corresponding SAR values

Three types of off-resonant binomial RF pulses were tested in this study: LIBOR, BORR and LIBRE. For LIBRE a total RF pulse duration of 1ms and 2.2ms was used, which are RF pulses that have been used in prior studies[4,6]. The RF excitation angle, the RF excitation frequency, the subpulses durations, and phase offset of each type of RF pulse is indicated. The tuning parameter, and its value yielding optimal fat suppression, is highlighted in orange.

Fat-free MRI with off-resonant LIBOR pulses – Mackowiak, Piccini, Bastiaansen

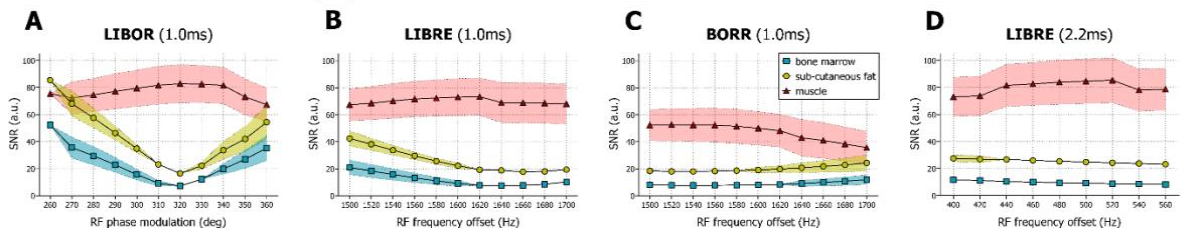

# Figure 3 – SNR of knee tissues in response to variation in RF frequency or phase excitation for the 4 water excitation pulses

The SNR in three ROIs located respectively in bone marrow, sub-cutaneous fat, and muscle (*vastus medialis*) was measured in 5 slices of a 3D acquisition using different off-resonant RF excitation pulses. The average and standard deviation across slices and volunteers (n=3) are plotted for each pulse as a function of its tuning parameter: RF phase modulation for LIBOR (A) and RF frequency offset for BORR (C) and LIBRE (B, D). An RF phase modulation of ~320° of the second LIBOR subpulse results in the highest fat signal suppression in subcutaneous fat and bone marrow, which is in agreement with a phase modulation of 315° obtained in numerical simulations (Fig.1). The drop in SNR of muscle tissue using the BORR pulse is most likely caused by an increase in background noise, which can be observed in the corresponding the knee images (Fig.4).

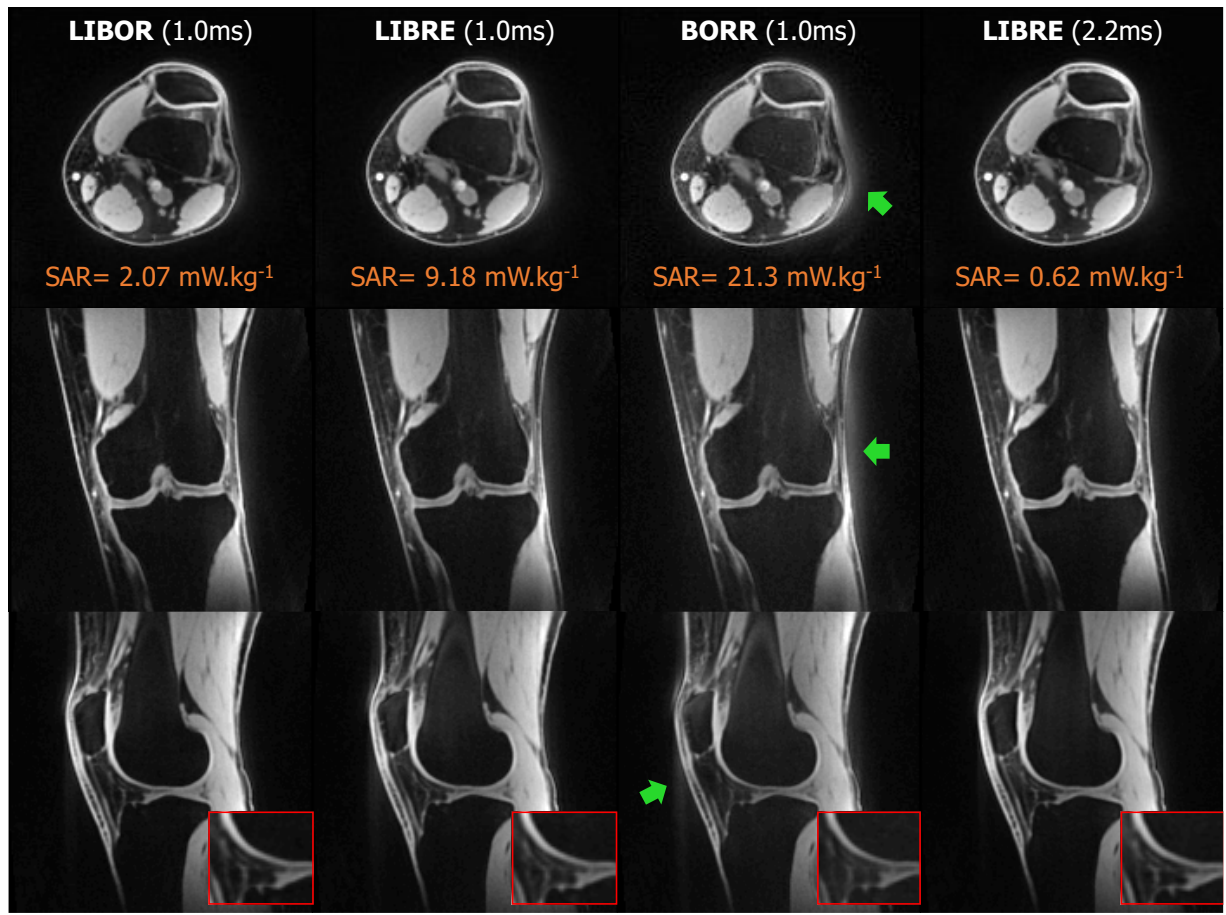

Figure 4 – Image comparison of binomial off-resonant RF pulses in knees

The fat-suppressed 3D MR images of the knee of research volunteer obtained with LIBOR, LIBRE (1ms), BORR, and LIBRE (2.2ms) are shown in transversal, coronal, and sagittal orientations. Although the suppression bandwidths were different for each pulse, the suppression of fat signal in the human knee was comparable. A close-up image including the lateral meniscus is shown. BORR images exhibit increased signal bleed in air-interfaced regions (green arrows).

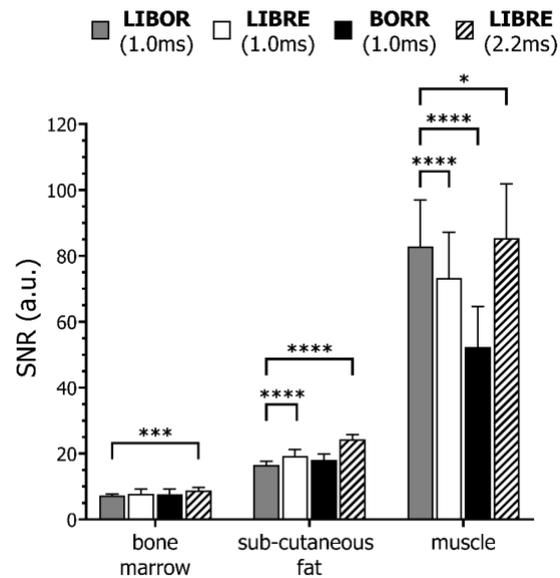

Figure 5 – SNR comparison of binomial off-resonant RF pulses in knees

The average SNR and standard deviation across subjects in three tissue types (bone marrow, sub-cutaneous fat, and muscle) was computed for each of the four pulses. Significant differences between LIBOR and the other water excitation pulses are indicated with asterisks (*). The drop in SNR of muscle tissue using the BORR pulse is likely caused by an increase of background noise, which is evident in the knee images in Fig.4 (green arrows). The increase in muscle SNR with the LIBRE (2.2ms) pulse may be related to the longer TR used in this experiment.

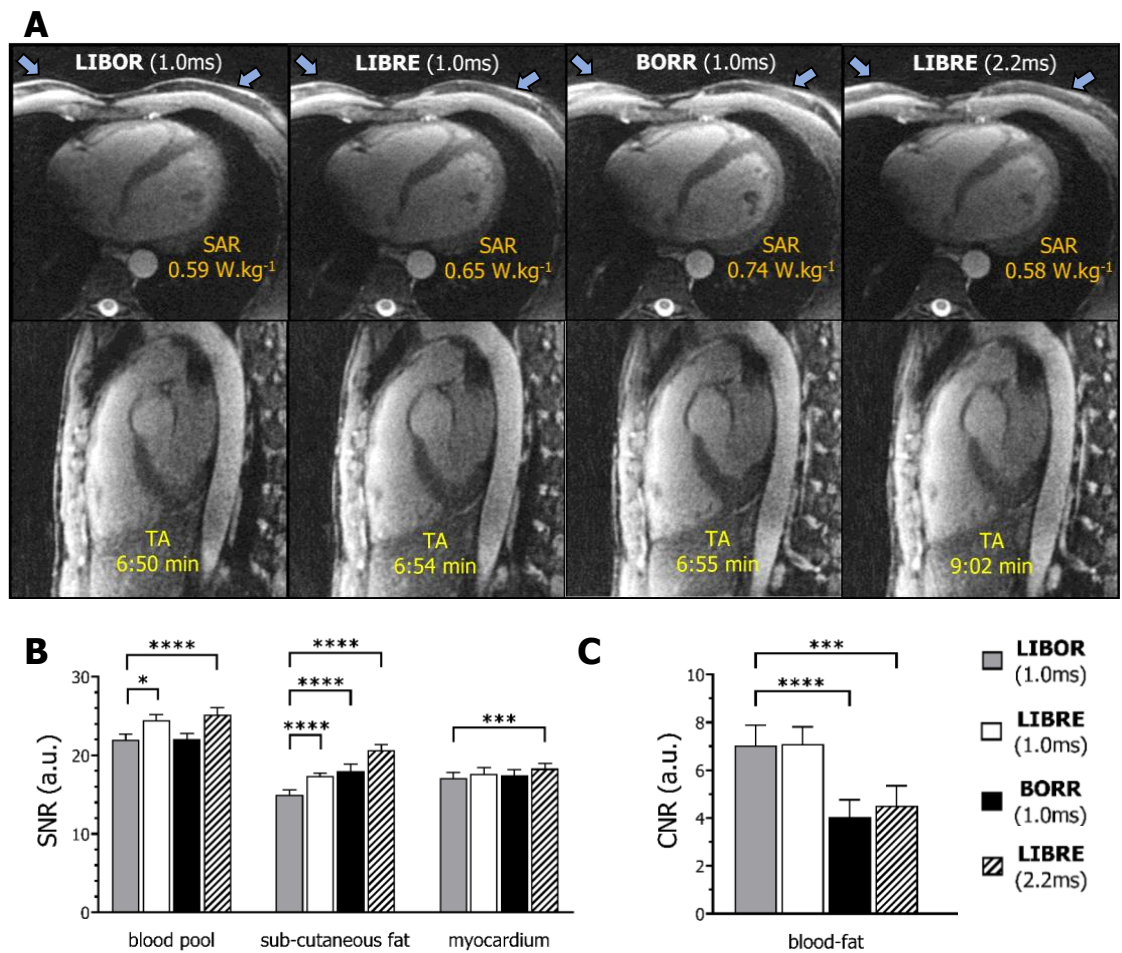

Figure 6 – Free-breathing respiratory self-navigated whole-heart MRI using different off-resonant water-excitation pulses

Transversal and sagittal views of 3D cardiac MR images in a volunteer, obtained with LIBOR, BORR, LIBRE (1ms) and LIBRE (2.2ms) are shown in (A). Fat signal suppression is nearly identical comparing the different RF pulses, but LIBOR maintains a short TR and simultaneously reduces SAR values. Slight variations in signal suppression of sub-cutaneous fat (A, blue arrows) could be observed, and statistically significant differences were found in SNR (B) and CNR (C) analysis. The acquisition times and SAR values are specified for each utilized RF pulse. SAR values are predominantly influenced by the power demanding T2 preparation module, which was used to generate contrast between ventricular blood and myocardium in the T1-weighted GRE sequence[25].

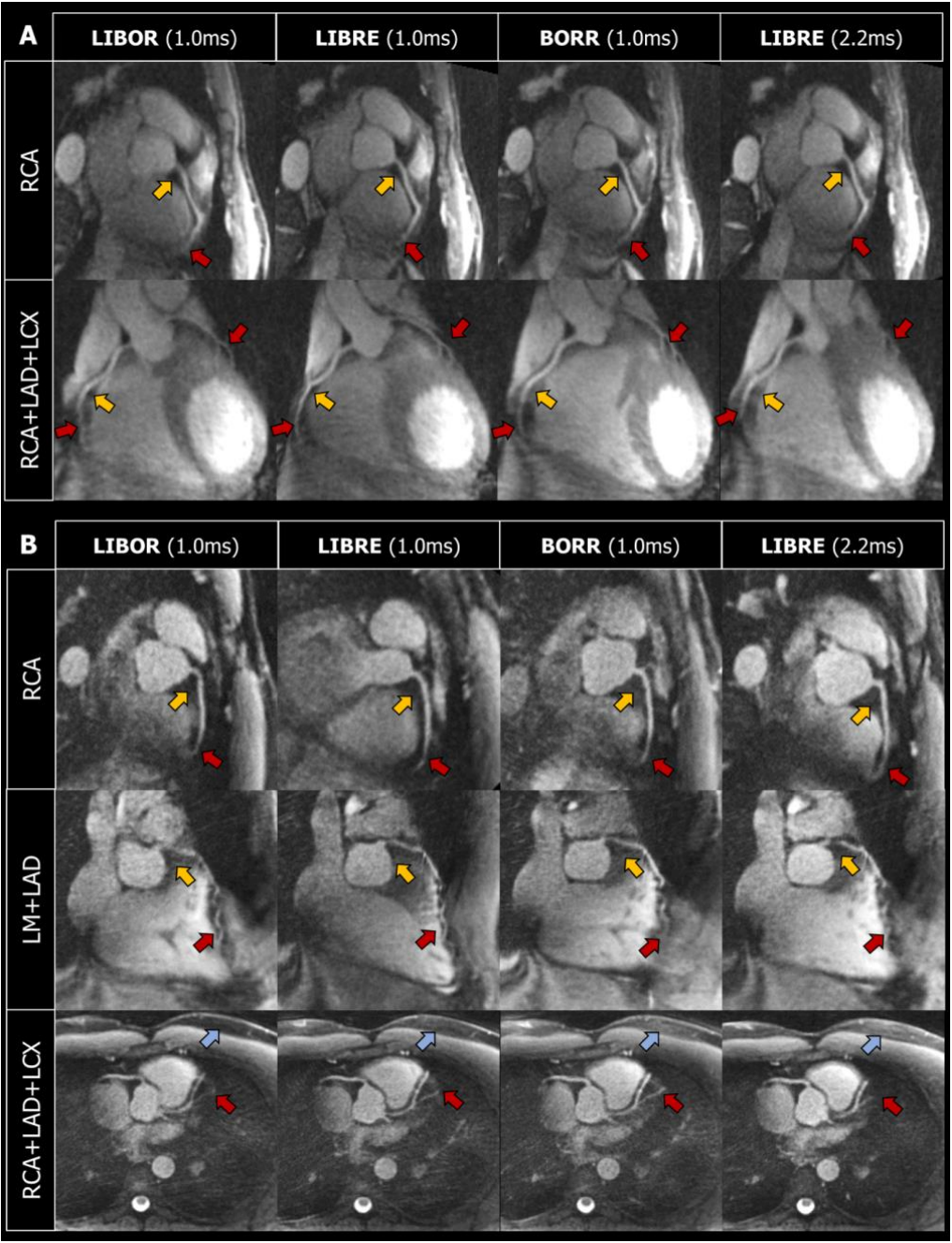

## Figure 7 – Coronary reformats in 2 volunteers

CMRA reformatted images from 2 volunteers (panels A and B) obtained with SoapBubble[28]. Transversal and coronal reformats were produced to visualize the right coronary artery (RCA), the left main (LM) and left anterior descending (LAD) arteries, as well as the left circumflex (LCX) coronary artery, when visible. Red arrows indicate (expected) coronary artery edges, yellow arrows indicate epicardial fat regions, and blue arrows indicate sub-cutaneous chest fat.

Fat-free MRI with off-resonant LIBOR pulses – Mackowiak, Piccini, Bastiaansen

|  | **LIBOR (1.0ms)** | **LIBRE (1.0ms)** | **BORR (1.0ms)** | **LIBRE (2.2ms)** |
|---|---|---|---|---|
| Average TA (min:s) | 6:49 ± 0:39 | 6:52 ± 0:38 | 6:51 ± 0:33 | 8:29 ± 0:39 |
| **RCA** | | | | |
| Detection rate | 5/5 | 5/5 | 5/5 | 5/5 |
| % vessel sharpness (first 4cm) | 36.2 ± 7.1 | 35.7 ± 9.5 | 34.3 ± 8.0 | 37.5 ± 8.4 |
| **LM+LAD** | | | | |
| Detection rate | 4/5 | 4/5 | 4/5 | 4/5 |
| % vessel sharpness (first 4cm) | 28.4 ± 5.4 | 29.7 ± 4.6 | 27.7 ± 5.1 | 27.0 ± 6.4 |
| **LCX** | | | | |
| Detection rate | 3/5 | 3/5 | 3/5 | 3/5 |

Table 2 – Detection and sharpness of cardiac vessels in volunteer experiments

Whole-heart free-breathing cardiac MRI performed using off-resonant LIBOR, BORR and LIBRE pulses were analysed for the detection of the right coronary artery (RCA), the left main (LM) and the left anterior descending (LAD) coronary artery, and the left circumflex (LCX) coronary artery. The percentage of vessel sharpness along the first 4cm of the vessel (when applicable) was measured and reported as an average and standard deviation across volunteers. Specific Absorption Rate (SAR) values and average scan time (TA) obtained with each pulse are indicated in the first two rows. While the 1ms-long RF pulses provided a significant scan time acceleration compared with the 2.2ms LIRBE pulse sequence, no significant differences in percentage of vessel sharpness were reported.